\documentclass[11pt,letterpaper]{article}

\usepackage[margin=1in]{geometry}
\usepackage{latexsym,graphicx,amssymb,amsbsy}
\usepackage{amsthm}
\usepackage{amsmath}
\usepackage{mathdots}
\usepackage{float}
\usepackage{xspace}
\usepackage{paralist}
\usepackage{enumerate}
\usepackage{cases}
\usepackage{caption}

\usepackage[ruled]{algorithm2e} 
\usepackage{algorithmic}

\usepackage{blkarray}

\usepackage{xcolor}
\usepackage{multicol}
\usepackage{graphicx}
\usepackage{subcaption}
\usepackage{varwidth}
\usepackage{wrapfig}
\usepackage{bbm}
\usepackage{tcolorbox}

\numberwithin{figure}{section}
\numberwithin{equation}{section}

\newtheorem{corollary}{Corollary}[section]
\newtheorem{theorem}{Theorem}[section]
\newtheorem*{theorem*}{Theorem}
\newtheorem{lemma}{Lemma}[section]
\newtheorem{claim}{Claim}[section]

\newcommand{\be}{\begin{equation}}
\newcommand{\ee}{\end{equation}}
\newcommand{\beq}{\begin{equation*}}
\newcommand{\eeq}{\end{equation*}}

\newcommand{\argmax}{\mathop{\rm argmax}}

\newcommand{\R}{\mathbb{R}}

\newcommand{\eps}{\varepsilon}
\newcommand{\ind}[1]{\mathbbm{1}\left[\vphantom{\sum}#1\right]}

\newcommand{\AutoAdjust}[3]{\mathchoice{ \left #1 #2  \right #3}{#1 #2 #3}{#1 #2 #3}{#1 #2 #3} }
\newcommand{\Xcomment}[1]{{}}

\newcommand{\InBrackets}[1]{\AutoAdjust{[}{#1}{]}}

\newcommand{\Exlong}[2][]{\operatornamewithlimits{\mathbf E}\limits_{#1}\InBrackets{#2}}
\newcommand{\Prx}[2][]{\operatorname{\mathbf{Pr}}_{#1}\InBrackets{#2}}

\newcommand{\eqdef}{\overset{\mathrm{def}}{=\mathrel{\mkern-3mu}=}}
\newcommand{\vect}[1]{\ensuremath{\mathbf{#1}}}

\newcommand\restr[2]{{
  \left.\kern-\nulldelimiterspace 
  #1 
  \vphantom{\big|} 
  \right|_{#2} 
  }}

\newcommand{\maxtwo}{\mathop{\rm max2}}

\renewcommand{\emptyset}{\varnothing}

\newcommand{\AP}[1][]{\textsf{AP}_{#1}} 
\newcommand{\AR}[1][]{\textsf{AR}_{#1}} 
\newcommand{\SPP}{\textsf{SPP}} 
\newcommand{\SPA}{\textsf{SPA}} 
\newcommand{\SPAvirtual}{\widetilde{\textsf{SPA}}} 

\newcommand{\APC}[1][]{\textsf{APC}_{#1}} 
\newcommand{\ARC}[1][]{\textsf{ARC}_{#1}} 

\newcommand{\MYER}{\textsf{Myerson}} 

\newcommand{\dd}{\mathrm{d}}

\newcommand{\N}{{\mathbb N}}

\newcommand{\prices}{\vect{p}}

\newcommand{\dist}{\mathbf{F}}
\newcommand{\dists}{\vect{\dist}}

\newcommand{\val}{v}
\newcommand{\vals}{\vect{\val}}
\newcommand{\valsmi}[1][i]{\vals_{\text{-}#1}}
\newcommand{\vali}[1][i]{{\val_{#1}}}

\newcommand{\costs}{\vect{c}}

\newcommand{\Tbar}{\overline{T}}

\begin{document}

%

\title{Bidder Subset Selection Problem in Auction Design
\thanks{This work is supported by Science and Technology Innovation 2030 –“New Generation of Artificial Intelligence” Major Project No.(2018AAA0100903), Innovation Program of Shanghai Municipal Education Commission, Program for Innovative Research Team of Shanghai University of Finance and Economics (IRTSHUFE) and the Fundamental Research Funds for the Central Universities. Zhihao Gavin Tang is supported by NSFC grant 61902233. Nick Gravin is supported by NSFC grant 62150610500.
} 
}

\author{
Xiaohui Bei\thanks{Nanyang Technological University, Email:  \texttt{xhbei@ntu.edu.sg}}
\and
Nick Gravin\thanks{ITCS, Shanghai University of Finance and Economics, Email: \texttt{\{nikolai, tang.zhihao,lu.pinyan\}@mail.shufe.edu.cn}}
\and
Pinyan Lu\footnotemark[2]
\and
Zhihao Gavin Tang\footnotemark[2]
}

\date{}
\maketitle

\begin{abstract}
Motivated by practical concerns in the online advertising industry, we study a bidder subset selection problem in single-item auctions. In this problem, a large pool of candidate bidders have independent values sampled from known prior distributions. The seller needs to pick a subset of bidders and run a given auction format on the selected subset to maximize her expected revenue. We propose two frameworks for the subset restrictions: (i) capacity constraint on the set of selected bidders; and (ii) incurred costs for the bidders invited to the auction. For the second-price auction with anonymous reserve (SPA-AR), we give constant approximation polynomial time algorithms in both frameworks (in the latter framework under mild assumptions about the market). Our results are in stark contrast to the previous work of Mehta, Nadav, Psomas, Rubinstein [NeurIPS 2020], who showed hardness of approximation for the SPA without a reserve price. We also give complimentary approximation results for other well-studied auction formats such as anonymous posted pricing and sequential posted pricing. On a technical level, we find that the revenue of SPA-AR as a set function $f(S)$ of its bidders $S$ is fractionally-subadditive but not submodular. Our bidder selection problem with invitation costs is a natural question about (approximately) answering a demand oracle for $f(\cdot)$ under a given vector of costs, a common computational assumption in the literature on combinatorial auctions.

\end{abstract}

\clearpage

\setcounter{page}{1}

\section{Introduction}
\label{sec:intro}

The competition among bidders and low dependency on the prior information\footnote{It is known as the ``Wilson doctrine'', named  after the leading advocate for the independence of mechanism design from the precise prior knowledge about the fine details of the environment.} are the main contributing factors to the prevalence of auctions as selling mechanisms. An exemplary result demonstrating the importance of competition in auctions is the seminal paper by Bullow and Klemperer~\cite{BK96}. It states that a commonly used second-price single-item auction with only one additional bidder extracts more revenue than the optimal auction for $n$ i.i.d. bidders. 
A more recent line of work on competition complexity initiated by Eden et al.~\cite{EFFTW17} (see also \cite{LiuP18,FeldmanFR18,BeyhaghiW19,CaiS21}) took the resource augmentation approach of~\cite{BK96} to the next level and considered the question of how many more additional bidders need to join the competition in order for a simple auction to surpass the revenue of the optimal mechanism in a variety of multi-item settings.

At a first glance it seems only beneficial for the auctioneer to have as many bidders entering the auction as possible, as the seller's revenue increases with more participants. 
However, this might not be the case in some real-world scenarios, as there could be inherent practical constraints on how many bidders can participate in the auction.
E.g., the auction's venue may have a fixed capacity that physically prevents too many participants, or there could be time-scheduling conflicts for the bidders to simultaneously participate in the auction. In addition, the seller may incur costs for inviting and accommodating bidders in the auction, which need to be factored in when computing auctioneer's profit.

The above concerns might appear to be moot for online and digital auctions conducted on the Internet, as the online communication is easy, fast, and cheap. But this is in fact far from truth, as we demonstrate with a concrete example of real-time bidding from the online advertising industry below. Note that online advertising is a prime example of auction design applied to modern economy and serves as a major source of revenue for many tech companies.

\paragraph{Real-time bidding.}
Most online ads are selected and displayed through \emph{real-time bidding (RTB)} -- each ad slot (an impression about the advertising opportunity) is sold in real-time by an automated auction process to a set of candidate advertisers. One might think that given the online nature of the RTB auctions, the seller could invite and serve as many advertisers in the auction as possible. However,
the issues we outlined above are even more pronounced in the RTB context.
\begin{itemize}
    \item 
    Many online advertising platforms already face (or are anticipating to face in a near future) a problem of too many advertisers participating in each online auction. 
    It is important to keep in mind that the entire process from initiating the auction to displaying the ad on the auctioned slot takes place in just under a few \emph{milliseconds}. Thus it is crucial for the platform to keep the communication lag and all necessary computations under a strict time limit, which is hard to achieve with a large number of bidders. The current practice is to keep this number under a threshold $m$ (for example, $m = 20$). That is, if there are more than $m$ bidders, the platform would need to select a subset of $m$ bidders first before running the auction.
    \item Secondly, in recent years more and more advertisers are delegating their businesses to professional digital marketing agencies (DMAs), who will interact with the advertising platform and perform bidding strategies on behalf of multiple advertisers.
    From the advertising platform's point of view, such practices reduce the number of participants and make bidder capacity less of an issue. However, the platform then needs to spend additional effort for inviting and maintaining business relations with these agencies, which would translate into costs that need to be subtracted from the auction revenue when computing the final profit of the platform.
\end{itemize}

In all the examples above, the auctioneer is facing a \emph{bidder selection} problem.
That is, how to select an optimal subset of bidders, subject to a capacity constraint or with costs, such that the expected revenue from running an auction with these bidders is maximized. This is the topic of our paper. 
More concretely, we focus on the setting of selling a single item to multiple bidders with independent values and consider a few concrete auction formats used in practice.

\subsection{Our Contributions}
We formulate a new theoretical framework for auction design with bidder selections.
More formally speaking, assume that there is a large pool $N$ of potential bidders, where each bidder $i\in N$ has a known prior distribution $F_i$. The platform selects first a set of bidders $S \subseteq N$, and then runs a specific auction format for the selected bidders. We mainly focus on the second price auction with an anonymous reserve price, which is perhaps the most relevant format in the context of online advertising. We also consider other types of auction formats, such as anonymous pricing, sequential posted pricing, and  Myerson auction. We consider the following two bidder selection models.

\paragraph{Bidder selection with capacity constraints.}
In the first model, the platform needs to select a subset $S$ of no more than $|S|\le m$ bidders and then run a fixed auction format to maximize its expected revenue. The auction format we are mostly interested in is the second price auction with an anonymous reserve price $r$ (the price can be optimized for the set $S$).

A closely related question was studied before by Mehta et al.~\cite{MehtaNPR20}. They consider the subset selection problem for the second price auction (i.e., the reserve price $r=0$) and show that it is NP-hard to approximate within a constant ratio. Note that no reserve price requirement is most natural for another problem of \emph{social welfare} maximization (also known as $k$-MAX problem), which has been shown first by Chen et al.~\cite{ChenHLLLL16} to admit a PTAS (see also~\cite{MehtaNPR20,SegevS21}). For the revenue objective, there is no practical restriction for the platform not to use a reserve price. We show that a subset of $m$ bidders and a reserve price $r$ can be found in 
polynomial time, so that the second price auction with reserve $r$ achieves constant approximation to the 
optimum solution. This is in stark contrast to \cite{MehtaNPR20} whose results indicate hardness of approximation of an apparently easier optimization problem, where the reserve price $r$ is set to be $0$. Our result easily follows from (i) an observation that the revenue of the second price auction always lies within a constant factor of the optimal revenue of the anonymous pricing; and (ii) that it is easy to compute the subset of agents that maximizes 
revenue of the anonymous pricing mechanism. 

In addition, for the sequential posted pricing mechanisms, we give a 2-approximation algorithm based on dynamic programming. 

\paragraph{Bidder selection with invitation costs.}
In the second model, each bidder $i$ has an \emph{invitation cost} $c_i$ that the platform needs to pay in order to include this bidder in the auction. The optimization goal of the platform is to select a subset $S \subseteq N$ of bidders with the maximum \emph{profit margin}, i.e., the expected revenue of the auction ran on the bidder set $S$ minus its total invitation cost $\sum_{i \in S}c_i$.

In this model, we find that the revenue of the the second price auction with anonymous reserve behaves as a fractionally subadditive function (a more general class of functions than submodular) over different subsets of bidders. The respective optimization problem with invitation costs becomes equivalent to giving an approximate solution to a demand oracle --- a commonly used theoretical abstraction in the literature on combinatorial auctions. We obtain a $O(1+\frac{1}{\delta})$-approximation for the difference objective under a mild assumption that the market is at least $\delta$-profitable, i.e., the revenue generated by the optimal solution is at least $(1+\delta)$ fraction of the cost of the optimal solution. The latter assumption certainly holds for the online advertising industry, usually with a rather large value of $\delta$. Our 
$O(1+\frac{1}{\delta})$-approximation result is much stronger than the typical bi-criteria approximation guarantees obtained for submodular maximization under costs\footnote{Bi-criteria results give two constant approximation guarantees when compared to the optimal solution: one for the value of the submodular function on the optimal set, another one for the total cost incurred by the optimal solution.} (see our discussion in the related works section).
 
We also obtain a 2-approximation result for the anonymous posted pricing mechanism. For the family of sequential posted pricing mechanisms, interestingly, our problem is closely related to the well studied pandora box problem. 

\begin{table}[h]
\centering
\begin{tabular}{|c|c|c|c|c|}
\hline
                    & $\AP$ & $\AR$ & $\SPP$ & $\MYER$ \\ \hline
Capacity Constraints & 1 & \begin{tabular}[c]{@{}c@{}} $1.64$ \\ Theorem~\ref{thm:AR_capacity} \end{tabular} & \begin{tabular}[c]{@{}c@{}} $2$ \\ Theorem~\ref{thm:spp}\end{tabular} & \begin{tabular}[c]{@{}c@{}} EPTAS \\ \cite{SegevS21,MehtaNPR20} \end{tabular} \\ \hline
Invitation Costs    & \begin{tabular}[c]{@{}c@{}} $2$ \\ Theorem~\ref{tm:alg_AP_knapsack} \end{tabular} & \begin{tabular}[c]{@{}c@{}} $O(1/\delta)$ \\ Theorem~\ref{th:delta_approx} \end{tabular} & - & - \\ \hline
\end{tabular}
\caption{Summary of Results: $\AP$ denotes the anonymous pricing mechanism. $\AR$ denotes the second price auction with anonymous reserve. $\SPP$ denotes the sequential posted pricing mechanism. $\MYER$ denotes the Myerson auction.}
\end{table}

\subsection{Related Works}
\label{sec:related}

\paragraph{$k$-MAX Selection.} A closely related setting to our work is to select $k$ out of $n$ random variables so that the expected maximum value is maximized. The problem has been studied multiple times under different contexts with different names by \cite{ChenHLLLL16, GoelGM10,SegevS21, MehtaNPR20}. Chen et al.~\cite{ChenHLLLL16} gave a PTAS for the probelm. Segev and Singla~\cite{SegevS21} and Mehta et al.~\cite{MehtaNPR20} improved the result to an EPTAS. As pointed out by Mehta et al.~\cite{MehtaNPR20}, the problem has applications for search engine~\cite{BroderCHSZ03}, procurement auction~\cite{Tan92,RongL07}, and team selection~\cite{KleinbergR18}. 
In Section~\ref{sec:prelim}, we observe another application to Myerson's auction with capacity constraints.


\paragraph{Sequential Posted Pricing with invitation costs.} We study sequential posted pricing with capacity constraints in Theorem~\ref{thm:spp}. For the setting with invitation costs, due to the online nature of sequential posted pricing mechanisms, a natural implementation of the mechanism is to approach the bidders one by one and stop once the item is sold. In other words, the set of invited bidders are adaptive, rather than a pre-selected set of bidders. This setting is out of the scope of our bidder selection with invitation costs model. On the other hand, it has been formalized as the online pandora's box problem by Esfandiari et al.~\cite{EsfandiariHLM19}. They reduced the problem to the prophet inequality setting and designed constant approximation algorithms.


\paragraph{Submodular maximization.} Our bidder selection problem is related to the line of work on submodular maximization, where the objective is $\max_{S:|S|\leq m}f(S)$ (in the capacity constraint setting) or $\max_S f(S)-c(S)$ (in the cost setting, here $c$ is a linear function) for a monotone submodular function $f$.
In the capacity constraint setting, a greedy algorithm can achieve $1-1/e$-approximation and this is the best possible~\cite{NWF78}. In the cost setting, although the objective function is still submodular, the greedy algorithm is not applicable as the objective function can have negative values.
Sviridenko~\cite{SviridenkoVW17} provide a $(1-1/e,1)$-approximation algorithm for this problem, i.e. the algorithm finds a set $S$ with $f(S)-c(S) \ge (1-1/e) \cdot f(S^*) - c(S^*)$, where $S^*$ is the optimal solution.
Note that if $f(s^*)(1-1/e)<c(S^*)$, then~\cite{SviridenkoVW17} does not give any approximation guarantees for the maximization problem $\max_{S}(f(S)-c(S))$ with costs. 
Furthermore, in general, there is no polynomial time algorithm with approximation ration better than $n^{1-\eps}$ for the difference objective\footnote{There is an easy APX-hardness reduction for coverage valuations with linear costs to the max-independent set in a graph G on $[n]$ vertices, which is a folklore result for the unconstrained submodular maximization.}.
Subsequently, Feldman~\cite{Feldman21} provided a guess-free algorithm and generalized the result to $(1-e^{-\beta},\beta)$-approximation for every $\beta \in [0,1]$. This is shown to be tight by Bodek and Feldman~\cite{BodekF22}. 
Recently, Lu et al.~\cite{LuYG21}, Sun et al.~\cite{SunXZW22}, and Bodek and Feldman~\cite{BodekF22} generalized the result to non-monotone submodular functions $f$ and also designed bi-criteria approximations.

We remark that true approximation results are not achievable in the cost setting for general submodular functions. Our theorem~\ref{tm:alg_AP_knapsack} on anonymous pricing mechanisms studies a specific family of XOS functions (a.k.a. fractionally subadditive, is a super-class of submodular functions) that allows positive results. For XOS function, in the capacity constraint setting, the problem does not admit a polynomial-time $m^{1/2-\epsilon}$-approximation algorithm in the value oracle model~\cite{singer2010budget}, but has a 2-approximation in the demand oracle model~\cite{badanidiyuru2012optimization}. 


\section{Preliminaries}
\label{sec:prelim}

We consider a single-item auction where the seller has one item to sell to a set of bidders.
Let $N$ denote the set of all potential bidders with $|N| = n$.
Each bidder $i\in N$ has a valuation $v_i$ for the item which is drawn from a prior distribution $F_i$ independently across all $n$ bidders. 
All prior distributions $\{F_i\}_{i \in N}$ are known to the seller. We use $f_i(p)$ and $F_i(p)=\Prx{v_i\le p}$ to denote respectively the 
probability and cumulative density functions of each distribution $F_i$. 

\paragraph{Auction with bidder selections.} The auction has two phases. In the first phase, the seller needs to select a subset of bidders $S \subseteq N$ to participate in the auction. In the second phase, the auction takes a vector of bidders' values $\vals = (v_i)_{i \in S}$ as the input, and decides which bidder (if any) receives the item and how much that bidder needs to pay.
In the setting with capacity constraint the auctioneer can only select a set $S: |S|\le m$ with at most $m$ bidders. In the setting with invitation costs, there is a known vector of costs $\costs=(c_i)_{i\in N}$ and no restrictions on the subset $S$.

\subsection{Auctions}
We will investigate the following types of auctions in this paper.
Below, the set $S \subseteq N$ indicates the set of selected bidders participating in the respective auction format.

\paragraph{Anonymous Pricing ($\AP[r]$):} the seller puts up a take-it-or-leave price $r$ to all bidders, and the item is sold to any bidder whose value for the item is no less than $r$.
Let $\AP[r](S)$ denote the expected revenue of the auction with the reserve price $r$ and bidders' set $S$. Then we have
\[\AP[r](S) = r \cdot [1 - \prod_{i \in S}F_i(r)].\]
We also use $\AP(S) = \max_{r}\AP[r](S)$ to denote the best revenue over all possible choices of $r$.

\paragraph{Second Price Auction with Anonymous Reserve ($\AR[r]$):} the seller sets a reserve price $r \ge 0$, and all bidders with values below this price are discarded;
the item is sold to the highest of the remaining bidders at a price equal to the second highest agent's value or the reserve price if none other remain.
We use $\AR[r](S)$ to denote the expected revenue of the auction with reserve price $r$ and bidders set $S$,
and use $\AR(S) = \max_r \AR[r](S)$ to denote the best revenue over all possible choices of $r$.

When the reserve price is $0$, this is the classic second-price auction. We denote its expected revenue as $\SPA(S)$.

\paragraph{Sequential Posted Pricing ($\SPP$):} it is defined by the bidder set $S$, price vector $\prices = (p_i)_{i \in S}$, and an order/permutation $\pi \in \Pi: S \mapsto [|S|]$ where $\pi(i)$ is the place of bidder $i$ in the arrival sequence for each $i \in S$; the bidders arrive one by one in the order $\pi$ (i.e., in the sequence $(\pi^{-1}(j))_{j\in[|S|]}$), and the item is sold to the first bidder $i$ with $v_i \geq p_i$.

We have three notations for the expected revenue of $\SPP$: (a) $\SPP(S, \prices, \pi)$ for given prices $\prices$ and order $\pi$; (b) when only the order $\pi$ is fixed, $\SPP(S, \pi) = \max_{\prices}\SPP(S, \prices, \pi)$ is the expected revenue of the auction with optimal prices; (c) when the seller is free to choose both the order and the prices, $\SPP(S) = \max_{\pi}\SPP(S, \pi)$.
For the ease of notation, we sometimes also write $\SPP(S, \prices, \pi)$ when $\pi: N \mapsto [n]$ is an order of all bidders in $N$. In this case we assume that bidders in $S$ arrive in the order $\pi$ restricted to $S$.

\paragraph{Myerson Auction ($\MYER$):} this is auction format that maximizes expected revenue~\cite{myerson1981optimal}; it 
considers the virtual value function $\phi_i(v) = v - \frac{1-F_i(v)}{f_i(v)}$ for each bidder $i \in N$;\footnote{If the distributions $(F_i)_{i\in N}$ are regular, i.e., each $\phi_i(v)$ is monotonically non decreasing in $v$. For irregular distributions one should use ironed virtual functions instead (see, e.g., \cite[Section 3.3.5]{hartline2013mechanism} for more details).}
the item is allocated to the bidder (if any) with the highest non-negative virtual value $\max \{\phi_i(v_i) \mid \phi_i(v_i)\geq 0, i \in S\}$, and this bidder is charged their threshold bid to win the auction.
We let $\MYER(S)$ to denote the expected revenue extracted from the bidder set $S$.

\subsection{Simple Observations}

\begin{claim}[Section 3.3.3~\cite{hartline2013mechanism}]
    \label{cl:Myerson_as_welfare}
    The expected revenue of the Myerson auction $\MYER(S)$ is the same as the expected welfare of the second price auction $\SPAvirtual(S)$ for the distributions $(\widetilde{F}_i)_{i\in S},$ where $\widetilde{F}_i: \widetilde{v}_i=\max\{\phi_i(v_i),0\}$ with $v_i\sim F_i$.
\end{claim}
This claim allows us to convert the problem about revenue maximization for the Myerson auction format to the welfare maximization problem for the second-price auction, which has been already studied before~\cite{ChenHLLLL16,MehtaNPR20,SegevS21}.

\begin{corollary}[\cite{ChenHLLLL16,MehtaNPR20,SegevS21}]
    Revenue of the Myerson auction $\MYER(S)$ as a set function of the subset of bidders $S\subseteq N$ is a monotone submodular function.
    Moreover, there is an efficient PTAS for solving $\max_{S\subseteq N: |S|\le m}\MYER(S)$.
\end{corollary}
For the Anonymous Pricing it is convenient to fix the price $r$ first and then analyse $\AP[r](S)$.
\begin{claim}
    \label{cl:ap_submodular}
    For a fixed price $r$, $\AP[r](S)$ is a monotone submodular set function of $S\subseteq N$.
\end{claim}
\begin{proof}
As we have $\AP[r](S)= r \cdot [1 - \prod_{i \in S}F_i(r)]$, the function $\AP[r](S)$ is monotone in $S$ and non-negative. Furthermore, the marginal contribution $\AP[r](S)-\AP[r](S\setminus\{j\}) = r\cdot (1-F_j(r))\cdot \prod_{i \in S\setminus\{j\}}F_i(r)$ of a bidder $j\in S$ is a decreasing function of $S$. Hence, $\AP[r](S)$ is submodular.
\end{proof}
Furthermore, it is easy to compute the optimal set $S^*=\argmax_{S\subseteq N,|S|=m}\AP[r](S)$. 
\begin{claim}
    \label{cl:ap_capacity}
    For a fixed price $r$, it is optimal to select $m$ bidders with the highest probability of sale in $\{\Prx{\vali\ge r}\}_{i\in N}$
    as the solution $S^*=\argmax_{S\subseteq N, |S|=m}\AP[r](S)$.
\end{claim}    
Another good property of the anonymous pricing $\AP(S)$ is that it is a constant approximation to the revenue of the second-price auction with anonymous reserve $\AR(S)$.
\begin{claim}[\cite{jin2020tight}]
    \label{cl:ap_vs_ar_tight}
    For any set of bidders $S\subseteq N$ we have
    $1.64\cdot\AP(S)\approx\frac{\pi^2}{6}\AP(S)\ge \AR(S)\ge \AP(S)$.
\end{claim}

\section{Capacity Constraint} 
\label{sec:capacity}
In this section we study the bidder selection problem with a capacity constraint. Specifically, we consider the second price auction with anonymous reserve $\AR$ and sequential posted pricing $\SPP$ formats.

\subsection{Second Price Auction with Anonymous Reserve}
We start by analyzing the second price auction with anonymous reserve $\AR$. 
%
Note than when we set the reserve price $r = 0$, $\AR[r]$ becomes the classic second price auction \SPA (without a reserve price), and our problem becomes equivalent to selecting $m$ out of $n$ random variables with the objective of maximizing the expected second-highest value. This question was studied in~\cite{MehtaNPR20} with the following conclusion.

\begin{theorem}[\cite{MehtaNPR20}]
Assuming the exponential time hypothesis or the planted clique hypothesis, there is no polynomial time algorithm that, given $n$ random variables $X_1,\ldots, X_n$, finds a subset of size $m$ whose expected second largest value is a constant factor of the optimal.
\end{theorem}

This means that the optimization problem $\max_{S:|S|\leq m}\SPA(S)$ is APX-hard. However, this result does not necessarily imply the hardness for the objective $\max_{S:|S|\leq m}\AR(S)$. In fact, as we will see below, when allowing a positive reserve price, our objective can be effectively approximated.

We start by analyzing the property of the set function $\AR(S)$. 
It is known that certain families of set functions, such as monotone submodular functions, allow for efficient approximation algorithms. 
As it turns out, our function $\AR(S)$ is not submodular, but
belongs to a more general class of \emph{fractionally subadditive} (also known as \emph{XOS}) functions.

\begin{claim}
    \label{cl:XOS_AR}
    Revenue of the second price auction with optimal anonymous reserve $\AR(S)$ as a set function of the subset of bidders $S\subseteq N$ may be not submodular, but is a monotone XOS function. Moreover, the additive coefficients $\{A_i\ge 0\}_{i\in S}$ in the XOS representation of $\AR(S)$ (i.e., $\AR(S)=\sum_{i\in S}A_i$ and $\forall T\subseteq N ~~\AR(T)\ge \sum_{i\in T\cap S}A_i$) can be defined as follows
    \begin{align}
        \label{eq:XOS_coefficients}
        A_i=\Exlong[\valsmi]{p_i\cdot (1-F_i(p_i))},\text{ where } p_i(\valsmi)=\max_{j\in S-\{i\}}(v_j,r^*)
        \text{ is a random posted price,}\\
        \text{and }~r^*=\argmax_{r\in \R_+}\AR[r](S) \text{ is the optimal reserve price in }\AR(S).\nonumber
    \end{align}
\end{claim}
\begin{proof}
    To see that $\AR(S)$ is not submodular, consider the following instance with $n=3$ identical bidders for some small $\eps>0$: \[
    \forall i\in[3]~~ F_i\eqdef \begin{cases}
    v=1 & \text{ with probability } 99/100\\
    v\sim \textup{Uni}[1,1+\eps] & \text{ with probability } 1/100
    \end{cases}.
    \]
Then for any set $S\subseteq[3]$ the optimal reserve price $r^*$ in $\AR(S)$ is $r^*=1$. The revenue for a single bidder $\AR[1](\{1\})=1$; for two bidders $\AR[1](\{1,2\})=\AR[1](\{1,3\})=1+\frac{1}{100^2}\cdot\frac{\eps}{3}$, as the payment exceeds $1$ only when both bidders have values larger than $1$ and the surplus (on top of $1$) is equal to the revenue of $\SPA$ for two identical and uniform $\textup{Uni}[0,\eps]$ bidders; for three bidders $\AR[1](\{1,2,3\})\ge 1+3\cdot\frac{99}{100^3}\cdot\frac{\eps}{3}$, as exactly $2$ out of $3$ bidders have their values larger than $1$ with probability $3\cdot\frac{99}{100^3}$. We get that the marginal contribution of $2$ with respect to the set $\{1\}$ is smaller than its marginal contribution with respect to the set $\{1,3\}$: $\AR(\{1,2\})-\AR(\{1\})=\frac{\eps}{3\cdot 100^2}<
\frac{\eps\cdot 197}{3\cdot 100^3}\le \AR(\{1,2,3\})-\AR(\{1,3\})
$. 

Next we show that $\AR(S)$ is an XOS function with the respective additive coefficients $\{A_i\}_{i\in S}$ defined in \eqref{eq:XOS_coefficients}. Let us also define 
\begin{equation}
    \label{eq:XOS_coefficients_vals}
    A_i(\vals)\eqdef\ind{v_i\ge \max_{j\in S\text{-}\{i\}}(v_j,r^*)}\max_{j\in S\text{-}\{i\}}(v_j,r^*),\quad \text{then}\quad A_i=\Exlong[\vals]{A_i(\vals)},
\end{equation}
where $r^*$ is the optimal reserve price in $\AR(S)$. Then we have the following (we use $\maxtwo_{i\in S}(a_i)$ to denote the second largest number in a set $\{a_i\}_{i\in S}$).
\begin{multline*}
    \AR[r^*](S)=\Exlong[\vals]{\maxtwo_{i\in S}(v_i,r^*)\cdot\ind{\exists i\in S:v_i\ge r^*}}=\\
    \Exlong[\vals]{\sum_{i\in S}\ind{v_i\ge \max_{j\in S\text{-}\{i\}}(v_j,r^*)}\cdot \max_{j\in S\text{-}\{i\}}(v_j,r^*)}=\Exlong[\vals]{\sum_{i\in S}A_i(\vals)}=\sum_{i\in S} A_i.
\end{multline*}
It remains to verify that $\AR(T)\ge \sum_{i\in T}A_i$ for all $T\subseteq S$. Let $r_o=\argmax_{r}\AR[r](T)$ be the optimal price for $\AR(T)$ and $\Tbar=S\setminus T$. We shall use a random price $r\sim D$ instead of $r_o$ given by 
\[
r\eqdef\max_{j\in \Tbar}(v_j,r^*) \text{ for a random } \vals\sim\dists.
\]
Then 
\begin{multline*}
    \AR(T)=\AR[r_o](T)\ge\Exlong[r\sim D]{\AR[r](T)}=
    \Exlong[r(\vals_{\Tbar})]{
      \Exlong[\vals_{T}]{\sum_{i\in T}  
      \ind{v_i\ge \max_{j\in T\text{-}\{i\}}(v_j,r)}\cdot \max_{j\in T \text{-}\{i\}}(v_j,r)}}=\\
    \sum_{i\in T}\Exlong[\vals]{
    \ind{v_i\ge \max_{j\in S \text{-}\{i\}}(v_j,r^*)}\cdot \max_{j\in S \text{-}\{i\}}(v_j,r^*)
    }=\sum_{i\in T}A_i.
\end{multline*}
This concludes the proof of Claim~\ref{cl:XOS_AR}.
\end{proof}

From an optimization point of view, this characterization does not immediately help with our approximation objective, as it is known that general monotone XOS functions are hard to approximate within a factor better than $m^{1/2}$ in the value oracle model~\cite{singer2010budget}~\footnote{Although, XOS functions do allow for efficient approximation in the \emph{demand oracle model}. But a demand oracle is not easy to implement in our setting. Section~\ref{sec:costs} has a more detailed discussion on this issue}.
In order to obtain an efficient approximation algorithm, we take a detour and use closely related \emph{Anonymous Pricing}.

\begin{theorem}
    \label{thm:AR_capacity}
    There is a polynomial time algorithm that finds a set $S^*\subset N, |S^*|=m$ and a reserve price $r^*$ such that $\frac{\pi^2}{6}\AR[r^*](S^*)\ge\max_{S\subset N, |S|=m}\AR(S)$. 
\end{theorem}
\begin{proof}
By Claim~\ref{cl:ap_capacity} we can find an optimal set $S^*\in\argmax_{S\subseteq N, |S|= m}\AP(S)$ and optimal anonymous price $r^*$ in polynomial time by trying all possible anonymous reserve prices $r$. By Claim~\ref{cl:ap_vs_ar_tight} we have 
\[
\frac{\pi^2}{6}\AR[r^*](S^*)\ge \frac{\pi^2}{6}\AP(S^*)=
\max_{S\subseteq N, |S|=m} \frac{\pi^2}{6}\AP(S)\ge
\max_{S\subseteq N, |S|=m} \AR(S),
\]
which concludes the proof of Theorem~\ref{thm:AR_capacity}.
\end{proof}

\subsection{Sequential Posted Pricing}
\label{sec:spp_capacity}

In this section we focus on the sequential posted pricing mechanisms ($\SPP$). 
Recall that for any subset $S$ of bidders, $\SPP(S) = \max_{\prices, \pi} {\SPP(S, \prices, \pi)}$ is the optimal revenue from running a $\SPP$ with these bidders. Our objective is to find $\max_{S:|S|\leq m}\SPP(S)$.

In the following we will show a 2-approximation algorithm for this objective. 
The algorithm is based on two observations:
(1) for any fixed bidder arrival order $\pi$, and for any subset of bidders $S$, the optimal revenue $\SPP(S, \pi) = \max_{\prices} {\SPP(S, \prices, \pi)}$ is a 2-approximation to $\SPP(S)$; (2) when $\pi$ is fixed, $\max_{S:|S|\leq m} \SPP(S, \pi)$ can be computed efficiently by a dynamic programming algorithm.

\begin{theorem}
\label{thm:spp}
There is an efficient algorithm that provides a 2-approximation to the sequential posted price with capacity constraint problem.
\end{theorem}

\begin{proof}
We first fix an arbitrary arriving order $\pi_0 = (1, 2, \ldots, n)$ of the bidders. 
In the following we will show a dynamic programming algorithm that computes $\max_{S:|S|\leq m}\SPP(S, \pi_0)$.
We let 
\[g(i, k) = \max_{S \subseteq [i,n], |S|=k}\SPP(S, \pi_0), \qquad \forall 1\leq i\leq n,~~ 1 \leq k \leq n-i+1\]
denote the optimal revenue with respect to the order $\pi_0$ when we are allowed to select $k$ bidders from the set $\{i, i+1, \ldots, n\}$.
The function $g(\cdot,\cdot)$ satisfies the following recurrent relation

\[g(i, k) = \max \left\{g(i+1,k),~~ \max_{p_i} \left\{ p_i\cdot(1-F_i(p_i)) + g(i+1, k-1)\cdot F_i(p_i)\vphantom{\sum}\right\} \right\}.
\]
Here we consider two options: first, not to make any price offer to $i$; second, offer price $p_i$ to $i$, in which case we get the expected revenue of $p_i\cdot(1-F_i(p_i))$ from $i$, and if $i$ does not  buy the item with probability $F_i(p_i)$, we still can make up to $k-1$ offers to the remaining $\{i+1,\ldots,n\}$ bidders. The optimal price $p_i$ can be found by scanning through the list of all possible prices $p_i$.\footnote{If $F_i$ is a continuous distribution supported on an interval $[A,B]$ ($0<A<B$) we can find an $(1+\eps)$ approximation to all $g(i,k)$ by discretizing the possible price $p_i$ as multiples of $\frac{\eps}{n}$).}
This recurrence allows us to compute each $g(i, k)$ in a backward order, and $g(1, m)$ is the solution to our objective.

The final piece of the proof is the fact that for any fixed set of bidders $S$ and any order $\pi_0$, we have
\[\SPP(S, \pi_0) = \max_{\prices} {\SPP(S, \prices, \pi_0)} \geq \frac12 \max_{\pi} \max_{\prices} {\SPP(S, \prices, \pi)} = \frac12 \SPP(S)\]
where the inequality is a well-known result which can be found in, e.g.,~\cite[Section 4.2.2]{hartline2013mechanism}.
\end{proof}

\section{Invitation Costs}
\label{sec:costs}
In this section we consider the setting with costs. Our main focus here will be on the second price with anonymous reserve auction format. But before that, we present a 2-approximation result for the anonymous pricing mechanisms. Let $\APC(S)\eqdef\max_r\AP[r](S)-\sum_{i\in S}c_i$ denote our objective function for any set of bidders $S\subseteq N$. Let $S_o$ be the optimal set $S_o=\argmax_{S}\APC(S)$. We show in the following 
subsection how to compute a set $S$ in polynomial time so that $2\APC(S)\ge \APC(S_o)$. 

\subsection{Anonymous Pricing}
\label{sec:ap_costs}
We consider the problem when the price $r$ of the anonymous pricing mechanism is fixed, i.e. we want to maximize $\APC[r](S) \eqdef \AP[r](S) - \sum_{i\in S}c_i$ for a given $r$. 
Our main result is a $2$-approximation polynomial-time algorithm for this objective.

With a fixed price $r$ every bidder $i$ has two important parameters: 1) the probability that $\Prx{v_i \ge r}$ and 2) the invitation cost $c_i$. Let $w_i \eqdef -\ln\left(1-\Prx{v_i \ge r}\right)$. Then $\AP[r](S) = r \cdot \Prx{\exists i \in S, v_i \ge r} = r \cdot \left( 1 - \exp\left(-\sum_{i \in S} w_i \right) \right)$. Our optimization problem can be written as follows.
\begin{equation}
\label{eq:ap_integral}
\max_{S\subseteq N} ~\APC[r](S)~~ = ~~\max_{S\subseteq N} :~ r \cdot \left( 1 - \exp\left(-\sum\nolimits_{i \in S} w_i \right) \right) - \sum\nolimits_{i \in S} c_i
\end{equation}
A simple observation is that we only need to consider those bidders $i$ with $\APC[r](\{i\}) \ge 0$. Formally, we prove the following.
\begin{lemma}
$\forall i \in S_o$,  we have $\APC[r](\{i\}) = r \cdot (1-\exp(-w_i)) - c_i \ge 0$.
\end{lemma}
\begin{proof}
By Claim~\ref{cl:ap_submodular} $\AP[r]$ is a submodular function. If there is a bidder $i\in S_o$ with $\APC[r](\{i\})<0$, then  $\APC[r](S_o)-\APC[r](S_o\setminus\{i\})=\AP[r](S_o)-\AP[r](S_o\setminus\{i\}) -c_i \le \AP[r](\{i\})-\AP[r](\emptyset)-c_i=\APC[r](\{i\})-\APC[r](\emptyset)<0$. I.e., $\APC[r](S_o)<\APC[r](S_o\setminus\{i\})$ -- a contradiction.
\end{proof}

From now on, we may assume without loss of generality that every bidder $i\in [n]$ satisfies $\APC[r](\{i\}) \ge 0$. 
We study the following natural fractional relaxation of \eqref{eq:ap_integral} similar to the fractional relaxation of the knapsack problem:
\begin{equation}
\label{eq:ap_frac}
\max_{\vect{x} \in [0,1]^n}: ~r \cdot \left( 1 - \exp\left(-\sum\nolimits_{i \in [n]} w_i \cdot x_i \right) \right) - \sum\nolimits_{i \in [n]} c_i \cdot x_i
\end{equation}
\paragraph{Integrality Gap.} We would like to remark that the integrality gap of the above optimization problem can be arbitrarily large even when there is only $1$ bidder. E.g., consider the function $h(x) = 1-e^{-w\cdot x} - (1-\eps) \cdot x$ for $w > 1$. Then when $w$ goes to infinity, $\max\{h(0),h(1)\} = \eps - e^{-w} \to \eps$, while $\max_{x \in [0,1]} h(x) \ge h(1/w) = 1 - 1/e - 1/w \to 1-1/e$.
On the positive side, we have the unbounded integrality gap only when there is a bidder $i$ with the corresponding \emph{non-monotone} in $x\in[0,1]$ function $h_i(x) = r \cdot \left( 1 - e^{-w_i \cdot x} \right) -c_i \cdot x$  (equivalently, $h'(1) = r \cdot e^{-w_i} \cdot w_i - c_i < 0$). We call such bidders \emph{special}.
Let $T \eqdef \{i \in [n] \mid r \cdot \exp(-w_i) \cdot w_i < c_i\}$ be the set of special bidders.
The next lemma shows that the optimal solution to \eqref{eq:ap_integral} has at most one special bidder.

\begin{lemma}
\label{lem:ap_large_bidder}
The optimal set of bidders satisfies that $|S_o \cap T| \le 1$.
\end{lemma}
\begin{proof}
Assume towards a contradiction that there are 
two different bidders $j, k \in S_o \cap T$. Then
\begin{equation}
\label{eq:derivative}
w_j \ge \exp(-w_k) \cdot (\exp(w_j) -1) \quad \text{or} \quad w_k \ge \exp(-w_j) \cdot (\exp(w_k) -1),
\end{equation}
as otherwise $w_j \cdot w_k < (1-\exp(-w_j)) \cdot (1-\exp(-w_j))$, contradicting the fact that $w_j > 1-\exp(-w_j)$ and $w_k > 1-\exp(-w_k)$. Without loss of generality,  we assume the first inequality of \eqref{eq:derivative} holds. Then, we have
\begin{multline*}
\APC[r](S_o \setminus j) = r \cdot \left( 1 - \exp\left(-\sum_{i \in S_o } w_i + w_j \right) \right) - \sum_{i \in S_o} c_i + c_j 
> r \cdot \left( 1 - \exp\left(-\sum_{i \in S_o } w_i + w_j \right) \right)\\ - \sum_{i \in S_o} c_i + r \cdot \exp(-w_j) \cdot w_j 
\ge r \cdot \left( 1 - \exp\left(-\sum_{i \in S_o } w_i + w_j \right) \right) - \sum_{i \in S_o} c_i + r \cdot \exp(-w_k) \cdot \left(1 - \exp(-w_j)\right) \\
\ge r \cdot \left( 1 - \exp\left(-\sum_{i \in S_o } w_i + w_j \right) \right) - \sum_{i \in S_o} c_i + r \cdot (1- \exp(-w_j)) \cdot \exp\left(-\sum_{i \in S_o} w_i +w_j\right) = \APC[r](S_o),
\end{multline*}
contradicting the optimality of $S_o$. Here, the first inequality holds as $j\in T$ and by the definition of $T$; the second inequality holds by \eqref{eq:derivative}.
\end{proof}

According to the above lemma, we can enumerate all possible choices of a bidder $t\in T\cap S_o$ and ignore all other bidders in $T$ in the fractional relaxation \eqref{eq:ap_frac}. This allows us to
get a slight variation of \eqref{eq:ap_frac} with a bounded integrality gap. Specifically, we consider 
the following fractional relaxation \eqref{eq:ap_frac2}, where the extra parameter $q\in[0,1]$ corresponds to the event of selecting the bidder $t$ in advance and the optimization is restricted to a subset of bidders $S\eqdef N\setminus T$
\begin{equation}
\label{eq:ap_frac2}
\textup{FR}(S,q): \qquad \max_{\vect{x} \in [0,1]^S}~ r \cdot \left( 1 - q \cdot \exp\left(-\sum_{i \in S} w_i \cdot x_i \right) \right) - \sum_{i \in S} c_i \cdot x_i
\end{equation}
Note that the objective function in \eqref{eq:ap_frac2} is concave in $\vect{x}$. By checking local maximum conditions for the optimal solution $\vect{x^*}$ to $\textup{FR}(S,q)$, it is easy to see that almost the same as the standard greedy algorithm for the fractional knapsack problem finds $\vect{x^*}$ (see the description of the algorithm below). Similar to the fractional knapsack problem there is at most one fractional bidder in $\vect{x^*}$.
\begin{enumerate}
    \item Sort the bidders in $S$ in the ascending order of $w_i/c_i$.
    \item For $i$ from $1$ to $|S|$: continuously increase $x_i$ until $x_i = 1$ or $r \cdot q \cdot w_i \cdot \exp(-\sum_{j \le i} w_i \cdot x_i) = c_i$.
\end{enumerate}

Finally, we provide our algorithm for approximating \eqref{eq:ap_integral}.
\begin{tcolorbox}
\begin{enumerate}
    \item For each $t \in T$, solve $\textup{FR}([n] \setminus T,\exp(-w_t))$ greedily and denote the optimal value and solution by $\textup{FR}^*(t)$ and $\vect{x}^*(t)$ respectively. Let $S^*(t) = \{i \mid x_i^*(t) = 1\} \cup \{t\}$.
    \item Solve $\textup{FR}([n] \setminus T, 1)$ greedily and denote the optimal value and solution by $\textup{FR}^*(0)$ and $\vect{x}^*(0)$ respectively. Let $S^*(0) = \{i \mid x_i^*(0) = 1\}$.
    \item Let $i^* = \argmax_i \APC[r](\{i\})$.
\end{enumerate}
Return the best solution among $\{i^*\}, S^*(0)$ and $S^*(t)$ for $t \in T$.
\end{tcolorbox}

\begin{theorem}
    \label{tm:alg_AP_knapsack}
    The above algorithm provides a $2$-approximation to the anonymous pricing with invitation costs problem.
\end{theorem}
\begin{proof}
For arbitrary $t \in T \cup \{0\}$, there exists at most $1$ item $i(t)$ with fractional value, i.e. $x_{i(t)} \in (0,1)$. Since $i(t) \in [n] \setminus T$, we must have
\begin{equation}
\label{eq:rounding_singleton}
c_{i(t)} \le r \cdot \exp(-w_{i(t)}) \cdot w_{i(t)} \le r \cdot \exp(-w_{i(t)}) \cdot \frac{\exp(w_{i(t)} \cdot (1- x^*_{i(t)})) - 1}{1-x^*_{i(t)}},
\end{equation}
where the second inequality holds since $e^x-1 \ge x$ for $x \ge 0$.
Consequently,
\begin{multline*}
\textup{FR}^*(t) - \APC[r](S^*(t)) = r \cdot \exp\left( -\sum\nolimits_{j \in S^*(t)} w_j \right) \cdot (1 - \exp(-w_{i(t)} \cdot x_{i(t)}^*)) - c_{i(t)} \cdot x_{i(t)}^* \\
\le r \cdot (1 - \exp(-w_{i(t)} \cdot x_{i(t)}^*)) - c_{i(t)} \cdot x_{i(t)}^*  \le r \cdot (1 - \exp(-w_{i(t)})) - c_{i(t)} = \APC[r](\{i(t)\}) \le \APC[r](\{i^*\}),
\end{multline*}
where the second inequality is equivalent to \eqref{eq:rounding_singleton}. Finally, we conclude the proof of the theorem by Lemma~\ref{lem:ap_large_bidder} and the above inequality:
$\APC[r](S_o) \le \max_{t \in T \cup \{0\}}\textup{FR}^*(t)$ and
\[
\max_{t \in T \cup \{0\}} \textup{FR}^*(t) \le \max_{t \in T \cup \{0\}} \left( \APC[r](S^*(t)) + \APC[r](i^*) \right) \le \APC[r](\{i^*\}) + \max_{t \in T\cup \{0\}} \APC[r](S^*(t))~.
\]
Which concludes the proof of the Theorem~\ref{tm:alg_AP_knapsack}.
\end{proof}

For completeness, we consider the decision version of the anonymous pricing with costs and prove its NP-hardness.
Given $n$ bidders each with value distribution $F_i$ and cost $c_i$, and a number $R$, the question is to decide whether there exists a subset $S$ of bidders such that $\APC(S) \ge R$.
\begin{theorem}
The decision version of anonymous pricing with costs problem is NP-hard.
\end{theorem}
\begin{proof}
We consider the subset-sum problem and reduce it to our problem. Given an instance $I=(\{w_1,w_2,...,w_n\},W)$ for the subset-sum problem, the question is to decide whether a subset of $\{w_1,...,w_n\}$ sums to exactly $W$. We construct the following instance for the anonymous pricing with invitation costs problem with $n$ bidders. The value distribution of each bidder $i$ is a Bernoulli random variable: $v_i = 1$ with probability $1-e^{-w_i}$ and $v_i=0$ otherwise. The invitation cost of bidder $i$ is $c_i = e^{-W} \cdot w_i$. It is easy to observe that for any subset $S$ of bidders, the optimal anonymous pricing mechanism should set the price to be $1$, i.e. 
\[
\APC(S) = \max_r \AP(S) - \sum_{i \in S} c_i = \AP[1](S) - \sum_{i \in S} c_i = 1 - e^{-\sum_{i \in S}w_i} - e^{-W} \cdot \sum_{i \in S}w_i, \quad \forall S \subseteq [n]~.
\]
Let $h(x) \eqdef 1 - e^{-x} - e^{-W} \cdot x$. Notice that $h'(x) = e^{-x} - e^{-W}$. The function $h(x)$ achieves its maximum value of $1-e^{-W}-e^{-W} \cdot W$ at $x=W$. Therefore, to solve the subset-sum instance $I$, it suffices to decide whether $\max_S g(S) \ge 1-e^{-W}-e^{-W}\cdot W$. Due to the NP-hardness of the subset-sum problem, we conclude the proof of the theorem.
\end{proof}

\subsection{Second Price auction with Anonymous Reserve.}
Here we study the problem of maximizing the revenue of the second price auction with anonymous reserve minus the invitation cost, i.e., our objective is 
\[
\max_{S\subseteq N}\ARC(S), \quad\quad\quad\text{where } \ARC(S)=\AR(S)-\sum_{i\in S}c_i.
\]  
We have already seen in Section~\ref{sec:capacity} that anonymous pricing $\max_r \AP[r](S^*)$ gives a computationally efficient proxy with good approximation to $\AR(S^*)$. This was good enough to achieve a constant approximation for the setting with capacity constraint, but the difference objective makes the problem much more challenging. Indeed, for a general submodular function $f$ (and our function $f$ belongs to a more general class of XOS functions) the best we can hope for is a bi-criteria approximation of $\AR(S)$ and $c(S)=\sum_{i\in S}c_i$. The difficulty of approximating the difference objective becomes quite apparent in the regime where $\AR(S)$ is only slightly larger than $\sum_{i\in S}c_i$, in which case any approximation $f(S^*)$ of $\AR(S^*)$ must be accurate up to $1+\eps$ factor. Thus the naive substitution of the function $\AR(S^*)$ with a constant approximation $\AP(S^*)$ does not work in the cost setting. Still there is a hope that $\APC(S)=\AP(S)-\sum_{i\in S}c_i$ might be useful in approximating $\ARC(S^*)$ for a different set $S\ne S^*$.

On the positive side, we give the following approximation result parameterised by $\delta$ when the optimal solution $\AR(S^*)$ has a $\delta$-surplus over the cost $c(S^*)$, i.e., when $\AR(S^*)\ge(1+\delta)\sum_{i\in S^*}c_i$.
\begin{theorem}
    \label{th:delta_approx}
    Let $S^*$ be the optimum solution to $\max_{S} \ARC(S)=\max_{S}\left(\AR(S)-\sum_{i\in S}c_i\right)$. Suppose $S^*$ satisfies a $\delta$-surplus condition $\AR(S^*)\ge(1+\delta)\sum_{i\in S^*}c_i$. 
    Then there is a set of bidders $S$ with $O\left(\frac{1+\delta}{\delta}\right)\APC(S)\ge\ARC(S^*)$.
\end{theorem}
\begin{proof}
We use the probabilistic method to show existence of such set $S$. The idea is to randomly partition set $S^*$ into $\ell=O(\frac{1}{\delta})$ disjoint groups and then show that the combined revenues of the best anonymous pricing for each set in the partition almost cover $\AR(S^*)$. Specifically, we assign each bidder $i\in S^*$ independently at random to one of the $[\ell]$ groups and will use a carefully chosen random price for each set in the random partition to obtain the desired bound. Let $r^*=\argmax_r\AR[r](S^*)$ be the optimal reserve price. We have the following lower bound on the sum of revenues for anonymous pricing applied to a random partition of $S^*$.
\begin{multline}
    \label{eq:random_partition}
    \Exlong[\substack{R_1,\ldots,R_{\ell}\\\bigsqcup_j R_j=S^*}]{\sum_{j=1}^{\ell}\AP(R_j)}\ge
    \Exlong[\substack{R_1,\ldots,R_{\ell}\\\bigsqcup_j R_j = S^*}] {\Exlong[\vals]{\sum_{j=1}^{\ell}
    \overbrace{\AP[r_j(\vals)](R_j)}^{r_j(\vals)=\max_{k\in S^*-R_j}(v_k,r^*)}
    }}=\\
    \Exlong[\substack{R_1,\ldots,R_{\ell}\\\bigsqcup_j R_j = S^*}]{\Exlong[\vals]{\sum_{i\in S^*}
    \sum_{j=1}^{\ell}\ind{i\in R_j}\cdot\ind{v_i\ge \max_{k\in S^*-R_j}(v_k,r^*)}\cdot
    \ind{i=\argmax_{k\in R_j} v_k}\cdot r_j(\vals)}}\ge\\
    \Exlong[\vals]{\sum_{i\in S^*}\ind{i=\argmax_{k\in S^*}v_k, v_i\ge r^*} \Exlong[\substack{R_1,\ldots,R_{\ell}\\ \bigsqcup_j R_j = S^*}]{\sum_{j=1}^{\ell}\ind{i\in R_j}\cdot r_j(\vals)
    }}\ge\\
    \Exlong[\vals]{\sum_{i\in S^*}\ind{i=\argmax_{k\in S^*}v_k, v_i\ge r^*}\cdot\frac{\ell-1}{\ell}\maxtwo_{k\in S^*}(v_k,r^*)}=    \frac{\ell-1}{\ell}\AR[r^*](S^*),
\end{multline}
where to get the first inequality we simply changed the optimal price of each set $R_j$ to a specific random price $r_j(\vals)=\max_{k\in S^*-R_j}(v_k,r^*)$; in the first equality we used that each $v_i$ in $\vals$ has the same distribution $F_i$ as in $\AP[r_j(\vals)]$, and also charged the price $r_j(\vals)$ to the top bidder $i\in R_j$ (we assume that $\argmax$ returns a single number, say the smallest in a lexicographic ordering, in case of a tie); 
to get the second inequality we rearranged the order of summations and expectations and observed that only the top bidder $i\in S^*$ can be charged the price $r_j(\vals)$ (in fact, this inequality is equality when the chance that more than one bidder has the top value is $0$); 
to get the third inequality we observe that with probability $\frac{\ell-1}{\ell}$ the top bidder $i\in R_j$ and the 2-nd highest bidder are assigned to different groups of the partition $(R_1,\ldots,R_\ell)$ (in this case $r_j(\vals)=\maxtwo_{k\in S^*}(v_k,r^*)$ is the same as the payment in the second price auction with anonymous reserve $r^*$ for the set of bidders $S^*$).

Now, we set $\ell=\lceil\frac{2(1+\delta)}{\delta}\rceil$. Then by \eqref{eq:random_partition} there must exists a partitioning of $S^*$ into $R_1,\ldots,R_{\ell}$ such that 
\begin{equation}
\label{eq:ap_cover_ar}
    \sum_{j=1}^{\ell}\AP(R_j)\ge \left(1-\frac{1}{\ell}\right )\AR[r^*](S^*)\ge \left(1-\frac{\delta}{2(1+\delta)}\right)\AP(S^*).
\end{equation}
Finally, we get the following by summing all $\APC(R_j)=\AP(R_j)-\sum_{i\in R_j}c_i$.
\begin{multline*}
\sum_{j=1}^{\ell}\APC(R_j)=\sum_{j=1}^{\ell}\AP(R_j)-\sum_{i\in S^*}c_i\ge\left(1-\frac{\delta}{2(1+\delta)}\right)\AR(S^*)-\sum_{i\in S^*}c_i\\
=\frac{1}{2}\left(\AR(S^*)-\sum_{i\in S^*}c_i\right)+\frac{1}{2}\cdot \frac{\AR(S^*)}{1+\delta}-\frac{1}{2}\cdot\sum_{i\in S^*}c_i\ge
\frac{1}{2}\left(\AR(S^*)-\sum_{i\in S^*}c_i\right)=\frac{\ARC(S^*)}{2},
\end{multline*}
where the first inequality follows from \eqref{eq:ap_cover_ar}; the second inequality holds, as $\AR(S^*)\ge(1+\delta)\sum_{i\in S^*}c_i$.
Therefore, for one of $j\in[\ell]$ we have $\APC(R_j)\ge\frac{\ARC(S^*)}{2\ell}=O(\frac{1+\delta}{\delta})\ARC(S^*)$. 
\end{proof}
Theorem~\ref{th:delta_approx} coupled with the computationally efficient method for finding a set $T$ with $2\cdot\APC(T)\ge\APC(S)$ from  Section~\ref{sec:ap_costs} gives us the following. 
\begin{corollary}
    \label{cor:ar_costs_polytime}
    Let $S^*$ be the optimum set of bidders for second price auction with anonymous reserve $S^*=\argmax_{S} \ARC(S)$. Suppose $S^*$ satisfies a $\delta$-surplus condition $\AR(S^*)\ge(1+\delta)\sum_{i\in S^*}c_i$. 
    Then there is polynomial time algorithm that finds a set of bidders $S$ and a price $r$ such that anonymous pricing for set $S$ gets a surplus $O\left(\frac{1+\delta}{\delta}\right)\APC(S)\ge \ARC(S^*)$.  
\end{corollary}

On the negative side, when the generated revenue  $\AR(S^*)=(1+\delta)\sum_{i\in S^*}c_i$ is close to the total cost, the anonymous pricing may not approximate well $\ARC(S^*)$ for any subset $S\subseteq N$.

\begin{theorem}
There exists a set of $n$ i.i.d. bidders, such that the optimal profit achieved by anonymous pricing mechanism is at most $O\left(\frac{1}{\sqrt{n}}\right)$ portion of the optimal profit achieved by second price auction with anonymous reserve. I.e.,
\[
\max_{S} \APC(S) \le O\left(\frac{1}{\sqrt{n}}\right) \cdot \max_{S} \ARC(S)
\]
\end{theorem}
\begin{proof}
We define $n$ auxiliary functions: $G_i(v) \eqdef \left( 1 - \frac{i+1}{v} \right)^{\frac{1}{i}} \cdot \ind{v \in [i+1, \infty)}$ $\forall i \in [n]$. Consider the following instance of $n$ i.i.d. bidders with valuation distributions $F(v) \eqdef \max_{i \in [n]} G_i(v)$ and invitation costs of $c=1$ per bidder. Note that the bidders are identical as they are i.i.d. and have the same cost. We slightly abuse notations and use $\AP(i),\AR(i)$ to denote the revenue of anonymous pricing and the revenue of second price auction with anonymous reserve for $i$ bidders, also use $\APC(i), \ARC(i)$ to denote the corresponding surplus for $i$ bidders respectively.

We first give an upper bound of $\AP(i)$:
\[
\AP(i) = r \cdot (1 - (F(r))^i) \le r \cdot (1-(G_i(r))^i) \le i+1~.
\]
Consequently, we have $\max_i \APC(i) = \max_i \left(\AP(i) - i \right) \le \max_{i} (i+1 -i) = 1$. We calculate next the revenue of the second price auction for $n$ bidders:
\[
\AR[0](n) = n \cdot \int_{v \ge 2} v \cdot (1-F(v))~ \dd F^{n-1}(v),
\]
where, we interpret the second price auction as $n$ disjoint posted price mechanisms to each of $n$ bidders; the price is random and is equal to the maximum among the other $n-1$ bids.
\begin{claim}
For $i \in \N$ and $v \ge i+1$, we have $\left(1 - \frac{1}{v} - \frac{1}{e \cdot v^{3/2}} \right)^{i} \ge 1 - \frac{i+1}{v}$.
\end{claim}
\begin{proof}
The inequality for $i=1$ is equivalent to $\frac{1}{v} \ge \frac{1}{e\cdot v^{3/2}}$ and holds trivially for $v \ge 2$. For $i \ge 2$, 
consider function $h(y) \eqdef (1-y)^i - \left(1-i \cdot y + \frac{i(i-1)}{2e^2} \cdot y^2 \right)$. Observe that for $y \in [0, \frac{2}{i}]$,
\begin{multline*}
h'(y) = -i \cdot (1-y)^{i-1} + i - \frac{i(i-1)}{e^2} y = iy \cdot \left( (1-y) + (1-y)^2 + \cdots + (1-y)^{i-2} - \frac{i-1}{e^2} \right) \\
\ge iy \cdot \left( (i-1) \cdot \left(1-\frac{2}{i} \right)^{i-2} - \frac{i-1}{e^2} \right) = i(i-1)y \cdot \left( \left(1-\frac{2}{i} \right)^{i-2} - \frac{1}{e^2} \right) \ge 0~.
\end{multline*}\
Consequently $h(y) \ge h(0) = 0$ for $y \in [0, \frac{2}{i}]$. Applying the inequality to $y = \frac{1}{v} + \frac{1}{e\cdot v^{3/2}}$, we have
\begin{multline*}
\left(1 - \frac{1}{v} - \frac{1}{e \cdot v^{3/2}} \right)^{i} \ge \left(1-i \cdot \left(\frac{1}{v} + \frac{1}{e \cdot v^{3/2}}\right) + \frac{i(i-1)}{2e^2} \cdot \left(\frac{1}{v} + \frac{1}{e \cdot v^{3/2}}\right)^2 \right) \\
\ge 1 - \frac{i}{v} - \frac{i}{e \cdot v^{3/2}} + \frac{i^2}{4e^2} \cdot \frac{1}{v^2} = 1 - \frac{i+1}{v} + \frac{1}{v} \left( \frac{i}{2e \cdot v^{1/2}} - 1 \right)^2 \ge 1 - \frac{i+1}{v}~.
\end{multline*}
This concludes the proof of the Claim.
\end{proof}
According to the claim, we have that
\[
v \cdot (1-F(v)) = v \cdot \left(1 - \max_{i \le v-1} G_i(v) \right)  \ge v \cdot \left( \frac{1}{v} + \frac{1}{e\cdot v^{3/2}} \right).
\]
Therefore,
\begin{multline*}
\AR[0](n) = n \cdot \int_{v \ge 2} v \cdot (1-F(v)) \dd F^{n-1}(v) \ge n \cdot \int_{v \ge 2} \left( 1 + \frac{1}{e\cdot v^{1/2}} \right)  \dd F^{n-1}(v) \\
\ge n \cdot \left( 1 + \frac{1}{e} \cdot \int_{v \in [2,n]} \frac{1}{\sqrt{n}} \dd F^{n-1}(v) \right) = n + \frac{\sqrt{n}}{e} \cdot F^{n-1}(n) \\
\ge n + \frac{\sqrt{n}}{e} \cdot \max_i (G_{i}(n))^{n-1} \cdot  \ge n + \frac{\sqrt{n}}{e} \cdot (1- \frac{\sqrt{n}+1}{n})^{\frac{1}{\sqrt{n}}} = n + \Omega(\sqrt{n}).
\end{multline*}
We conclude the proof of the theorem by noticing that $\max_{i} \ARC(i) \ge \ARC(n) \ge \AR[0](n) - n \ge \Omega(\sqrt{n})$, while $\max_i \APC(i) = 1$.
\end{proof}

\section{Conclusion}
\label{sec:conclusion}

In this work we proposed two theoretical models for studying a practical problem of bidder selection in various simple auction formats. We obtained good approximation results for anonymous pricing, and found that anonymous pricing serves as a good proxy for other auction formats such as second-price auction with anonymous reserve.
We obtained generally good theoretical understanding for the model with capacity constraint.
However, there are still a number of open questions left in the model with invitation costs: a) if there is a PTAS, or APX hardness for anonymous posted pricing;
b) whether it is possible to get a constant approximation for the $\max_S \ARC(S)$ without large profit margin assumption; c) whether it is possible to obtain a constant approximation for the second price auction with the welfare objective (or revenue for the Myerson auction). 

There are a few future research direction that might be useful for practical applications. To name a few, i) study bi-criteria optimization trade-offs between the revenue and welfare objectives; ii) we assumed that bidders' distributions are perfectly known to the seller, which is not always true in practice. Analyse sample complexity of the subset selection problem, or study the problem in an online learning framework.
iii) From an engineering view point it is useful to have advice guided by theory about various heuristics algorithms. One approach that seem reasonable for the bidder selection problem and might be interesting from a theoretical point of view is local search algorithm. How fast it converges? What approximation guarantees can we have for the local optima?

\bibliographystyle{alpha}
\bibliography{bib}

\end{document}